\documentclass{iau} 
\usepackage{graphicx}
\usepackage{amsmath}	
\usepackage{amssymb}
\usepackage{siunitx}
\usepackage{caption}
\usepackage{floatrow}
\usepackage{color}
\usepackage{soul}
\usepackage[dvipsnames]{xcolor}

\definecolor{vero}{rgb}{0.5, 0.3, 0.7}
\definecolor{christi}{rgb}{0.0, 0.58, 0.71}
\definecolor{alexdu}{rgb}{0.00, 0.50, 0.00}
\definecolor{matt}{rgb}{0.8, 0.0, 0.0}

\newcommand{\sifourfull}{Si~\textsc{iv} $\lambda\lambda$1393, 1402}
\newcommand{\cfourfull}{C~\textsc{iv} $\lambda\lambda$1548, 1550}
\newcommand{\nfivefull}{N~\textsc{v} $\lambda\lambda$1238, 1242}

\newcommand{\stokesv}{Stokes $V$}

\title[Zeeman effect in Massive Star Magnetospheres] 
{Detecting the Zeeman effect in \\ Massive Star Magnetospheres in the UV}

\author[C. Erba, V. Petit, K. Gayley, R. Ignace, A. ud-Doula, G.A. Wade ]   
{C. Erba$^1$, 
V. Petit$^1$,
K. Gayley$^2$,
R. Ignace$^3$, \\
A. ud-Doula$^4$,
G.A. Wade$^5$
}

\affiliation{
$^1$Department of Physics and Astronomy, Bartol Research Institute, University of Delaware, Newark, DE 19716; email: cerba@udel.edu
\\[\affilskip]
$^2$Department of Physics and Astronomy, University of Iowa, Physics and Astronomy, Iowa City, IA, 52245
\\[\affilskip]
$^3$Department of Physics and Astronomy, East Tennessee State University, Johnson City, TN 37614
\\[\affilskip]
$^4$Department of Physics, Penn State Scranton, 120 Ridge View Drive, Dunmore, PA 18512
\\[\affilskip]
$^5$Department of Physics and Space Science, Royal Military College of Canada, PO Box 17000 Station Forces, Kingston, ON, Canada K7K 0C6
}
\pubyear{2021}
\volume{360}  
\jname{Astronomical Polarimetry 2020: New Era of Multiwavelength \\ Polarimetry}
\editors{H.Shinnaga, B-G Andersson, A.M.Magalh\~{a}es  \& E.Falgarone, eds.}

\begin{document}

\maketitle

\begin{abstract}
Approximately 7\% of massive stars host stable surface magnetic fields that are strong enough to alter stellar evolution through their effect on the stellar wind. It is therefore crucial to characterize the strength and structure of these large-scale fields in order to quantify their influence on massive star evolution. This is traditionally done by measuring the circular polarization caused by Zeeman splitting in optical photospheric lines, but we investigate here the possibility of detecting Stokes $V$ signatures in the wind-sensitive resonance lines formed in magnetically confined winds in the high opacity ultraviolet (UV) domain. This unique diagnostic would be accessible to high-sensitivity spaceborne UV spectropolarimeters such as \textsc{POLSTAR}.
\end{abstract}


\noindent \textbf{Introduction:} Approximately 7\% of massive stars have strong, stable, oblique, nearly dipolar magnetic fields with surface field strength typically $\sim$1 kG (Morel et al. 2015, Wade et al. 2016). These fields channel the stellar wind into a dynamic and structurally complex {\em magnetosphere}, altering stellar mass loss and rotation rates when compared to a non-magnetic star with similar spectral type (e.g. ud-Doula et al. 2009).

Magnetic channeling of the stellar wind can be studied by examining a star’s ultraviolet (UV) spectrum. Wind-sensitive UV resonance lines (e.g. \cfourfull, \sifourfull, and \nfivefull) provide key diagnostics of the kinematics, physical properties, and structure of massive star winds (e.g. Garcia \& Bianchi 2004). The details of the spectra are highly dependent on the strength of the magnetic field, the viewing angle of the star (orientation of the magnetosphere with respect to the observer), and the specific spectral lines observed. 

Zeeman splitting is present (but difficult to detect) in the spectral lines of magnetic massive stars (e.g. Donati \& Landstreet 2009). The components of the split line are circularly polarized. Circular polarization is detected and measured using \stokesv $= I_{L} - I_{R}$ profiles. Here, we explore the possibility of measuring magnetospheric polarization using UV wind lines (e.g. Gayley \& Ignace 2010, Gayley 2017).

\begin{figure}
\centering 
\floatbox[{\capbeside\thisfloatsetup{capbesideposition={right,top},capbesidewidth=4.5cm}}]{figure}[\FBwidth]
{\caption{Synthetic intensity (top row) and Stokes~$V$ (bottom row) profiles created using the UV-ADM code (Erba et al. 2021) for representative stars with a small magnetosphere ($B_\mathrm{p}\sim$1~kG; red dashed lines) and a large magnetosphere ($B_\mathrm{p}\sim$16~kG; blue solid lines), for a pole-on (left column) and equator-on (right column) view of the magnetosphere.}\label{fig:stokesv}}
{\includegraphics[width=0.6\textwidth]{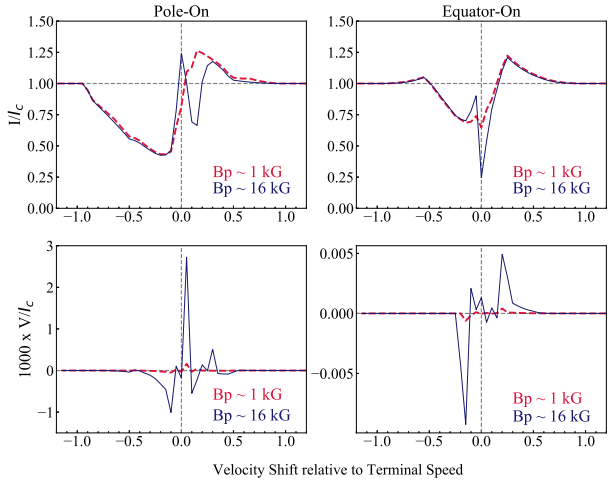} }
\end{figure}

\noindent \textbf{Constructing synthetic profiles using the UV-ADM Code:} Our goal is to model spectral lines in order to develop a more complete picture of the effects of global magnetic fields on massive stars. Since multidimensional magnetohydrodynamic (MHD) simulations are computationally expensive, we develop our models using the Analytic Dynamical Magnetosphere formalism (ADM; Owocki et al. 2016), which provides an analytic, time-averaged snapshot of the density, temperature, and velocity fields within the magnetosphere, and consequently dramatically diminishes computation time. We apply the ADM formalism to the UV regime, which is the region most sensitive to the kinematics of stellar outflows. We simultaneously treat the 3 regions (upflow wind, downflow wind, and hot post-shock region) of the ADM formalism, using a computationally efficient method to solve the equation of radiative transfer in 3D space, assuming a pure dipole field and using an approximation for the source function in the optically thin regime. Using the UV-ADM code (Erba et al. 2021), we have calculated a grid of synthetic UV (intensity) line profiles and \stokesv~profiles. Figure \ref{fig:stokesv} shows four representative models from that grid, demonstrating the difficulty of detecting \stokesv~signatures in stars with $B_\mathrm{p} \leq 1$~kG.

\noindent \textbf{Future Work:} We anticipate our model grid can be compared with UV data as a first step toward developing an understanding of the effects of strong magnetism in massive OB stars. Our work can be used to guide observational planning for upcoming missions like \textsc{polstar}, a proposed next-generation spaceborne 60-cm telescope and high-resolution UV spectropolarimeter with a mid-2028 launch date. \textsc{polstar}’s mission objectives include investigating the magnetically relevant phenomena that effect stellar and galactic evolution, such as wind inhomogeneities and stellar mass and angular momentum loss. 
Our results (Figure \ref{fig:stokesv}) indicate Zeeman splitting can inform these objectives for strongly magnetic targets at a $V/I$ sensitivity level of about 0.01\%.


\noindent \textbf{References:} \\
Donati \& Landstreet 2009, ARA\&A, 47, 333 \\
Erba et al. 2021, MNRAS, submitted \\
Garcia \& Bianchi 2004, ApJ, 606 497 \\
Gayley \& Ignace 2010, ApJ, 708, 615 \\
Gayley 2017, ApJ, 851, 113 \\
Morel et al. 2015, IAUS, 307, 342 \\
Owocki et al. 2016, MNRAS, 462, 3830 \\
ud-Doula et al. 2009, MNRAS 392, 1022 \\
Wade et al. 2016, MNRAS 456, 2 \\



\end{document}